# Beyond Permissions: A Configuration-Aware Empirical Assessment of Privacy Exposure in Children-Oriented and General-Audience Mobile Gaming Apps

Bakheet Aljedaani
Computer Science Department, Jamoum University College, Umm Al-Qura University, Makkah, Saudi Arabia
bhjedaani@uqu.edu.sa

## Abstract

Mobile gaming applications (apps) are among the most widely used mobile software systems, including a rapidly growing segment designed for children. These apps increasingly integrate analytics, advertising, and attribution infrastructures, which may introduce privacy and security exposure. Existing research has primarily examined permission usage or tracking behaviors in isolation, providing only a partial understanding of privacy exposure in modern mobile ecosystems. This study presents a configuration-aware, software engineering–oriented assessment of privacy exposure in mobile gaming apps by examining how permission usage, configuration-level design decisions, and third-party Software Development Kit (SDK) integrations jointly influence privacy exposure across children-oriented and general-audience apps. A systematic empirical study was conducted on 41 widely deployed Android mobile gaming apps using a three-phase methodology: (i) study protocol design, (ii) APK collection and static inspection, and (iii) data analysis. The analysis integrates permission inspection with manifest-level configuration indicators (e.g., backup settings, cleartext traffic, exported components) and third-party SDK ecosystem analysis. The results show that although children-oriented games generally request fewer permissions, they often exhibit comparable levels of privacy exposure due to configuration-level design choices and the integration of third-party tracking infrastructures. Across both categories, multiple apps demonstrated co-occurring exposure indicators, including permissive network configurations, exported components, and embedded analytics or advertising SDKs. The findings demonstrate that privacy exposure is primarily shaped by architectural and configuration-level decisions rather than permission requests alone. This study contributes a unified empirical assessment approach and provides actionable insights to support privacy-by-design in mobile software systems.

## 1. Introduction

Mobile applications (apps) have become integral to everyday life, supporting a wide range of domains including healthcare, education, entertainment, and social interaction. Among these, mobile gaming applications represent one of the most widely adopted categories, particularly among children and adolescents. In fact, the mobile gaming category is projected to dominate global app download volumes, with 34.54 billion app downloads across the App Store and Google Play combined, making it the largest app category by a wide margin [1]. Analysts predict that the mobile gaming market will grow to over $256 billion by 2030, driven by increasing smartphone adoption and growing engagement with digital games across diverse regions [2]. Mobile gaming apps primarily provide entertainment-oriented experiences to a broad user base; however, their large user bases and extensive integration with external services raise important security and privacy concerns. These concerns are amplified by the widespread use of app permissions, persistent network connectivity, third-party libraries, and data-driven monetization mechanisms such as advertising and analytics frameworks [3-5].

Several studies have highlighted privacy and security risks embedded within modern mobile apps, particularly those associated with mobile ecosystems that rely heavily on advertising, analytics, and attribution infrastructures. Large-scale empirical studies have shown that many mobile apps integrate complex third-party tracking libraries capable of collecting device identifiers, behavioral data, and usage telemetry without explicit user awareness [6-8]. These third-party components often operate independently of an application's visible functionality and may communicate with multiple external services, thereby expanding the potential data exposure surface of mobile apps. In addition to tracking infrastructures, software configuration decisions made during application development can introduce additional security and privacy exposures that remain largely invisible to end users. Prior studies have shown that Android app manifests frequently contain misconfigurations related to exported components, backup mechanisms, and network communication policies, which may increase the attack surface or expose application data under certain conditions [9, 10].

These concerns are particularly relevant in the context of mobile gaming applications, which represent one of the most widely installed categories of mobile apps worldwide and are frequently used by younger audiences. Mobile games commonly incorporate advertising and analytics frameworks to support monetization models, resulting in extensive integration with external services and data collection infrastructures [5, 11]. At the same time, many mobile games are designed for or widely used by children and adolescents, raising additional concerns regarding privacy exposure and data collection practices. Prior research in digital privacy has shown that younger users often have limited awareness of data collection practices and may be less capable of evaluating privacy exposures associated with mobile apps [3, 12]. Furthermore, studies suggest that permission requests alone provide only a partial view of privacy exposure, as significant data collection may occur through embedded third-party components and configuration-level design decisions (e.g., advertising or analytics SDKs) that operate beyond user-visible controls [6, 13, 14].

Despite growing attention to privacy exposure in mobile apps, existing empirical studies predominantly examine individual dimensions such as permission usage, user awareness, or third-party tracking behaviors in isolation. However, privacy exposure in modern mobile ecosystems is often shaped by the interaction between multiple architectural factors, including configuration-level design decisions and embedded third-party SDK ecosystems. As a result, current approaches provide only a partial and potentially misleading understanding of how privacy exposure emerges in practice. In particular, there is limited empirical evidence that systematically integrates permission analysis, configuration-level indicators, and third-party ecosystem dependencies into a unified assessment framework. Furthermore, comparative studies examining whether children-oriented mobile gaming apps are technically more privacy-preserving than general-audience apps remain scarce. To address this gap, this study adopts a configuration-aware, application-centric perspective that systematically evaluates how permissions, configuration-level attributes, and third-party SDK ecosystems jointly shape privacy exposure. Specifically, the study addresses the following research questions (RQs):

***RQ-1:*** *Do children-oriented mobile gaming apps request a different set of access permissions compared to general-audience mobile gaming apps?*

This RQ aims to compare the permission footprints of children-oriented and general-audience mobile games, focusing on differences in normal and sensitive permission usage.

***RQ-2:*** *To what extent do permission requests alone reflect the actual privacy exposure of children-oriented and general-audience mobile gaming apps?*

This RQ examines whether permission declarations, when considered in isolation, provide sufficient insight into privacy exposure, or whether additional configuration-level and architectural factors must be considered.

***RQ-3:*** *What configuration-level security and privacy exposures (e.g., backup exposure, cleartext traffic, exported components) are present in these apps?*

This RQ investigates manifest-level configuration and architectural choices that may introduce security and privacy exposures independent of permission requests.

***RQ-4:*** *How do third-party tracking and monetization ecosystems differ between children-oriented and general-audience mobile gaming apps?*

This RQ analyzes the integration of third-party analytics, advertising, and attribution SDKs and assesses how ecosystem dependencies influence privacy exposure across app categories.

To answer these research questions, we conducted a static analysis of a targeted dataset of widely installed Android mobile gaming apps, including both children-oriented and general-audience titles. We conducted an investigation involving 41 mobile gaming apps, which were categorized into two distinct groups: children-oriented gaming apps (comprising 21 apps) and general gaming apps (encompassing 20 apps). The apps were selected based on popularity indicators (e.g., installation counts and store rankings) to ensure that the analyzed apps represent widely used games with large user bases. The analysis was conducted in two complementary phases. First, we extracted and examined AndroidManifest.xml files to analyze permission requests, application configurations, and security-relevant settings (e.g., backup policies, network security configurations, and exported components). Second, we identified embedded third-party analytics, advertising, and attribution frameworks through static library inspection. The extracted data were then systematically synthesized and comparatively analyzed across app categories. The results showed that children-oriented games often request fewer permissions than general-audience games; however, permission minimization does not necessarily imply lower privacy exposure. Configuration-level exposures, including enabled cleartext traffic, unrestricted backups, and the presence of third-party tracking SDKs, were observed across both categories. These findings demonstrate that privacy exposure in mobile gaming apps is shaped not only by permission requests but also by architectural and configuration decisions embedded within the app ecosystem. The contributions of this study are fourfold:

- A configuration-aware empirical assessment approach for analyzing privacy exposure in mobile apps, integrating permission usage, manifest-level configuration indicators, and third-party SDK ecosystem analysis into a unified framework.
- A large-scale comparative empirical study of 41 widely deployed mobile gaming apps, providing systematic evidence on how configuration-level design choices and ecosystem dependencies influence privacy exposure beyond permission declarations.
- New insights into children-oriented mobile apps, demonstrating that reduced permission usage does not necessarily correspond to lower privacy exposure due to underlying architectural decisions and embedded third-party tracking infrastructures.
- Actionable software engineering implications, highlighting the role of configuration management and SDK integration practices in shaping privacy exposure and informing privacy-by-design strategies for mobile app development.

The remainder of this paper is organized as follows. Section 2 reviews related work. Section 3 details the research methodology. Section 4 presents the findings, followed by a discussion in Section 5. Threats to validity are discussed in Section 6, and Section 7 concludes the paper.

## 2. Related Work

Research on mobile app privacy and security has explored multiple dimensions of exposure in modern mobile ecosystems. Prior studies have examined Android permission usage, third-party tracking infrastructures, configuration-level security vulnerabilities, and privacy exposures affecting applications targeting children and young users. Together, these studies demonstrate that privacy exposure in mobile apps, in general, often arises from the interaction of several architectural factors rather than a single observable indicator. Hereafter, we review prior studies most relevant to our work and highlights the gap addressed by our study.

### *2.1 Privacy and Security in Android Apps*

A substantial body of research has investigated the effectiveness of Android's permission model as a mechanism for protecting user privacy. Empirical studies by Felt et al. in [15, 16] examined how users interact with Android permission prompts and found that many users do not fully understand the implications of permission requests. Their findings suggest that permission-based exposure communication alone may not adequately inform users about potential privacy exposures. Subsequent research further highlighted limitations of permission-based privacy assessment. For example, Reardon et al. in [17] demonstrated that Android apps can indirectly access sensitive information through shared resources, side channels, and interactions with embedded libraries, even when such access is not immediately visible through permission declarations. These findings indicate that permissions alone provide only a partial view of how apps collect and transmit user data. More studies continue to reinforce this observation. For example, Zimmeck et al. in [18] conducted a large-scale analysis of Android privacy policies and permissions and found significant inconsistencies between app behavior and disclosed data practices. Similarly, Li et al. in [19] performed a systematic literature review of Android security research and showed that static analysis techniques can reveal a wide range of security and privacy issues embedded within app architecture and configuration. Together, these studies demonstrate that evaluating privacy exposures in Android apps requires examining additional architectural elements beyond permission declarations.

### *2.2 Third-Party Tracking and SDK Ecosystems*

Another important stream of research focuses on the role of third-party libraries and SDKs in shaping privacy exposure within mobile apps. Mobile apps frequently integrate advertising, analytics, and attribution SDKs to support monetization and user engagement measurement. Early work by Grace et al. in [20] showed that embedded advertising libraries can access sensitive system resources and may introduce privacy and security risks independent of the host application's core functionality. Their analysis demonstrated that third-party libraries can significantly expand the data collection capabilities of mobile apps. More recent research has continued to investigate the influence of SDK ecosystems. For example, Laperdrix et al. in [5] analyzed tracking infrastructures in Android mobile games and showed that monetization strategies strongly influence the prevalence of embedded trackers, with free games typically integrating more third-party libraries than paid apps. Large-scale empirical studies have also examined SDK ecosystems more broadly. Meng et al. in [21] conducted a large-scale privacy assessment of Android third-party SDKs and identified widespread privacy-sensitive data flows as well as inconsistencies between observed behaviors and policy disclosures. Similarly, Rodriguez et al. in [22] examined privacy configuration options in widely used SDKs and found that default SDK settings often enable additional data collection features unless explicitly configured by developers. These findings suggest that third-party SDK ecosystems play a central role in shaping privacy exposure in modern mobile apps.

*2.3 Configuration-Level Security Exposures in Android Applications*

In addition to permissions and third-party libraries, configuration-level design decisions can also introduce privacy and security issues. Android app manifests define key security properties such as permission declarations, component exposure, backup policies, and network communication settings. A study by Chin et al. in [23] demonstrated that insecure inter-application communication and improperly exposed components can create exploitable conditions in Android apps. These issues arise from configuration choices rather than runtime permissions. Similarly, Alkinoon et al. in [24] showed that configuration-level design decisions, including exported components and network settings, can expose sensitive application data even when permission requests appear benign. Empirical studies in other application domains also highlight the importance of configuration-level security analysis. An empirical study by Forsberg and Iwaya in [25] investigated ten widely-used Android health and fitness apps using a combination of static and dynamic security analyses. The study uncovered multiple vulnerabilities, including insecure coding practices, hardcoded sensitive information, over-privileged permissions, misconfigurations, and extensive communication with third-party domains. Although the focus is on mHealth fitness apps rather than mobile games, this work highlights the prevalence of architectural and third-party exposure exposures and underscores the importance of assessing applications beyond permission sets. Another empirical study by Chen et al. in [26] examined the security posture of mobile banking apps through systematic analysis techniques, including static inspection of app components and configurations. The results revealed issues such as insecure data storage, improper cryptographic usage, excessive permissions, and vulnerabilities in backend communication, emphasizing that critical security and privacy issues often originate from architectural design choices and misconfigurations. Despite the different app domain, this study reinforces the need for comprehensive analysis beyond permission requests, particularly considering implementation decisions and third-party integrations.

*2.4 Privacy Issues in Children-Oriented Mobile Apps*

Recent research has also examined privacy issues in apps targeting children. Children-oriented mobile apps often rely on similar advertising and analytics infrastructures as general-audience apps, raising concerns regarding compliance with child data protection regulations. For example, Alomar and Egelman in [27] investigated how developers of child-directed mobile app approach privacy compliance and found that many developers struggle to interpret regulatory requirements and platform policies. More recent work by Alomar et al. in [28] examined the impact of app-store policy changes on privacy compliance and found that while policy updates improved transparency in some cases, privacy exposures associated with third-party SDK integration often remained. Saeed et al. in [29] showed that many users struggle to interpret privacy-related information provided by mobile apps, which further complicates users' ability to assess privacy exposures. Together, these studies highlight the need for technical analysis of privacy exposure in mobile apps, particularly in applications that may be widely used by children.

*Conclusive Summary:* Although prior research has examined Android permission models, third-party SDK ecosystems, configuration-level security risks, and privacy challenges in children-oriented apps, these studies largely focus on individual dimensions of privacy exposure in isolation. Few studies have systematically integrated these dimensions into a unified empirical framework that captures how privacy exposure emerges from the interaction of architectural factors. Moreover, comparative evidence focusing specifically on children-oriented and general-audience mobile gaming apps remains limited. To address these gaps, this study adopts a configuration-aware, application-centric approach that integrates permission analysis, manifest-level configuration inspection, and third-party SDK ecosystem analysis into a unified empirical assessment. By systematically comparing children-oriented and general-audience mobile gaming apps, this work provides new insights into how architectural design decisions and ecosystem dependencies jointly shape privacy exposure in mobile gaming environments.

## 3. Research Methodology

This study adopts a configuration-aware empirical assessment approach to investigate privacy exposure in Android mobile gaming applications. Rather than evaluating privacy exposures through a single dimension, the approach integrates three complementary perspectives: permission usage, configuration-level security indicators, and third-party SDK ecosystem dependencies. This multi-dimensional perspective enables a more comprehensive assessment of privacy exposure as an emergent property of app architecture. Static analysis is employed to systematically examine observable artifacts embedded within application packages, including permissions, manifest configurations, and third-party library integrations. The methodology is organized into three main phases: (1) study protocol design, (2) APK collection and static inspection, and (3) data synthesis and analysis, each detailed below. Figure 1 summarizes the end-to-end workflow adopted in this study, from defining the inspection criteria and collecting APKs to synthesizing static analysis results into qualitative privacy-exposure categories.

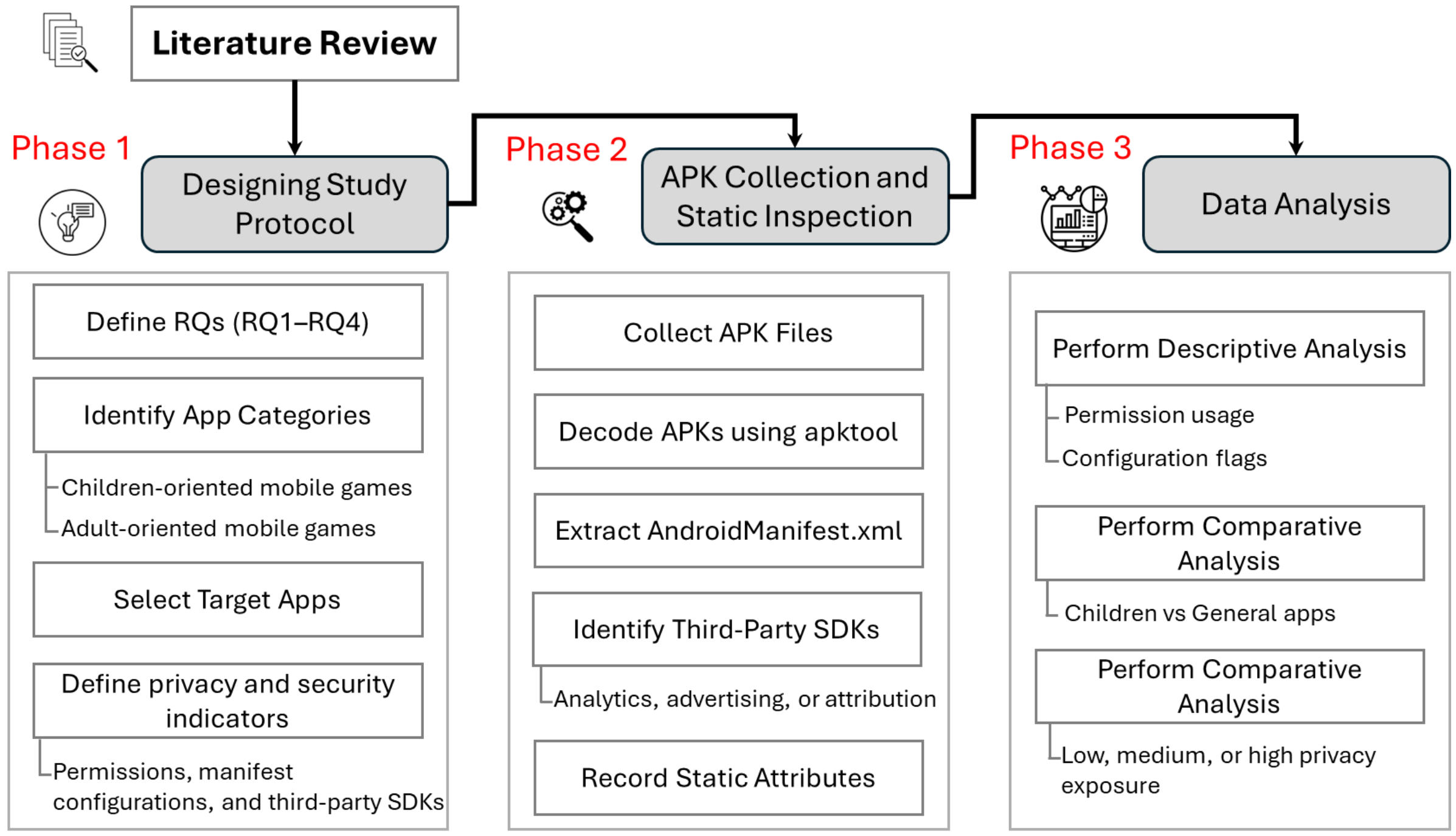

Figure 1. Overview of the Study Workflow

### *3.1 Phase 1 – Designing the Study Protocol*

The first phase focused on designing a structured study protocol to guide the empirical analysis. Based on the research gap identified in the literature, four research questions (**RQ1**–**RQ4**) were formulated to examine differences in privacy exposure indicators between children-oriented mobile games and general-audience mobile games. Then, the criteria for selecting apps were defined. Apps were selected according to the following criteria:

- The app belongs to the mobile gaming category in the Google Play Store.
- The app has high installation counts and popularity indicators, ensuring that the dataset reflects widely used games with large user bases.
- The app clearly targets either children-oriented audiences or general audiences, based on store descriptions, age ratings, and developer-provided metadata.
- The app package (APK) can be obtained and analyzed through static inspection.

Following these criteria, a dataset of 41 Android mobile gaming apps was constructed, including both children-oriented (20 apps) and general-audience (21 apps) games. The selected apps represent widely installed games across different genres and development frameworks. To support responsible reporting, apps are referred to using anonymized identifiers (e.g., **App1–App41**). The mapping between identifiers and original app names is provided in the Appendix. Finally, a set of privacy and security indicators was defined to guide the analysis, which were derived from prior research on Android apps security and privacy exposures, including studies on manifest misconfigurations, third-party tracking ecosystems, and mobile privacy leakage [6, 15, 17, 30]. These indicators include:

- Access permission requests, including both normal and sensitive Android permissions.
- Configuration-level attributes extracted from AndroidManifest.xml, such as backup policies, network security settings, and exported application components.

- The presence of embedded third-party analytics, advertising, and attribution SDKs, which may introduce additional data collection infrastructures.

These indicators allow the study to examine privacy exposure from multiple architectural perspectives rather than relying solely on permission analysis.

*3.2 Phase 2 - APK Collection and Static Inspection*

In the second phase, APK files corresponding to the selected apps were collected from publicly available repositories that mirror Google Play distributions. Each APK file was decoded using the Apktool framework, which enables extraction of application resources and configuration artifacts. Static inspection focused primarily on artifacts contained within the AndroidManifest.xml file, which defines permissions, component configurations, and security-related application settings. The static inspection process was conducted in two complementary steps:

*(i) Manifest-Level Analysis:* The manifest-level analysis was performed to extract key configuration attributes, including:

- Permission declarations.
- Backup configuration (*android:allowBackup*).
- Network security configuration settings.
- Component exposure (exported activities, services, and broadcast receivers).
- Cleartext network communication policies.

These attributes represent configuration decisions that may influence application security and privacy exposure [23].

*(ii) Third-Party SDK Identification:* The presence of third-party libraries was analyzed through static library inspection. SDK detection was performed by examining package names, class structures, and library signatures commonly associated with analytics, advertising, and attribution frameworks. This approach enables identification of third-party components embedded within the application even when they are not explicitly visible through permission declarations [20, 21].

All extracted attributes were recorded in a structured dataset to support consistent comparison across applications and categories (Available in Appendix).

*3.3 Phase 3 - Data Analysis*

The final phase focused on synthesizing and analyzing the extracted data. First, all collected indicators were consolidated into a structured dataset to enable systematic inspection and comparison. Descriptive statistics were then used to summarize permission usage, configuration attributes, and SDK prevalence across the analyzed applications. Followed by a comparative analysis which was conducted to identify similarities and differences between children-oriented and general-audience mobile games. This comparison allows the study to examine whether different app categories exhibit distinct privacy exposure patterns. To support interpretability of the results, each application was assigned a qualitative privacy exposure level based on the combined presence of relevant indicators. The categorization follows the following logic:

- *High exposure*: Multiple strong indicators co-occur, such as enabled cleartext communication together with embedded analytics or advertising infrastructures.
- *Medium exposure*: Partial exposure conditions are present, such as tracking indicators without cleartext communication or limited configuration exposures.
- *Low exposure*: Minimal indicators of configuration exposure or third-party tracking infrastructures are observed.

Importantly, this classification reflects potential privacy exposure inferred from static indicators, rather than confirmed misuse or exploitation of user data.

*Ethics and Responsible Disclosure*: This study relies exclusively on static inspection of publicly distributed Android app packages. The research does not involve exploiting app vulnerabilities, bypassing access controls, or collecting personal user data. All results are reported at an aggregate configuration level, focusing on permissions, manifest attributes, and third-party SDK indicators. No application-specific vulnerabilities or user-identifying information are disclosed. This approach aligns with responsible security research practices and minimizes potential harm to developers and end users.

## 4. Findings

This section presents the observed security and privacy indicators based on the conducted static analysis on 41 Android mobile gaming apps, including 20 children-oriented games (48.8%) and 21 general-audience games (51.2%). As illustrated in Section 3, the analysis examined several indicators associated with potential privacy and security exposure, including permission usage, configuration-level attributes, network communication settings, and the integration of third-party SDKs. All the shown code snippets were extracted from the decoded AndroidManifest.xml files of the investigated apps using static analysis and are shown as representative examples. For illustrative reasons, we provide visual markings, wherever required, to express a conclusive summary of the subsection. It is important to emphasize that the results presented in this section represent observable indicators derived from static analysis, rather than confirmed exploitation or malicious behavior.

### *4.1 Permission Usage and Sensitive Capabilities*

Permission requests represent one of the primary mechanisms through which Android apps access device resources. Previous research has shown that excessive or unnecessary permission requests may increase the potential exposure of user data [15, 31]. Across the dataset, all analyzed mobile games declared multiple permissions in their Android manifests, reflecting baseline requirements such as network connectivity and basic device interaction. Children-oriented apps requested an average of 11.35 permissions per application, whereas general-audience games requested an average of 17.43 permissions, corresponding to approximately 54% more permissions on average. As illustrated in Figure 2, network-related permissions such as INTERNET and ACCESS_NETWORK_STATE are nearly ubiquitous across both categories. However, sensitive permission usage is more prominent among general-audience games, suggesting that permission counts alone provide limited insight into actual privacy exposure.

In several cases, children-oriented apps declared multiple permissions while avoiding sensitive capabilities, whereas some general-audience games with relatively modest permission counts still integrated tracking or monetization infrastructures. These observations reinforce the notion that permission declarations alone provide an incomplete representation of privacy exposure. Overall, 28 of the 41 analyzed apps (68.3%) requested at least one permission categorized as sensitive within the dataset. This included 11 of the 20 children-oriented apps (55%) and 17 of the 21 general-audience apps (81%). For example, App2, a general-audience game, requested multiple sensitive permissions and was categorized as a high privacy exposure app in the dataset. Similarly, App11 and App14 requested several permissions extending the scope of device resource access. In contrast, App1, a children-oriented game, requested a more limited set of permissions and was categorized as a medium exposure app. Additionally, App24 and App29, both children-oriented game, include permissions associated with external storage access, indicating functionality related to cached content or media handling.

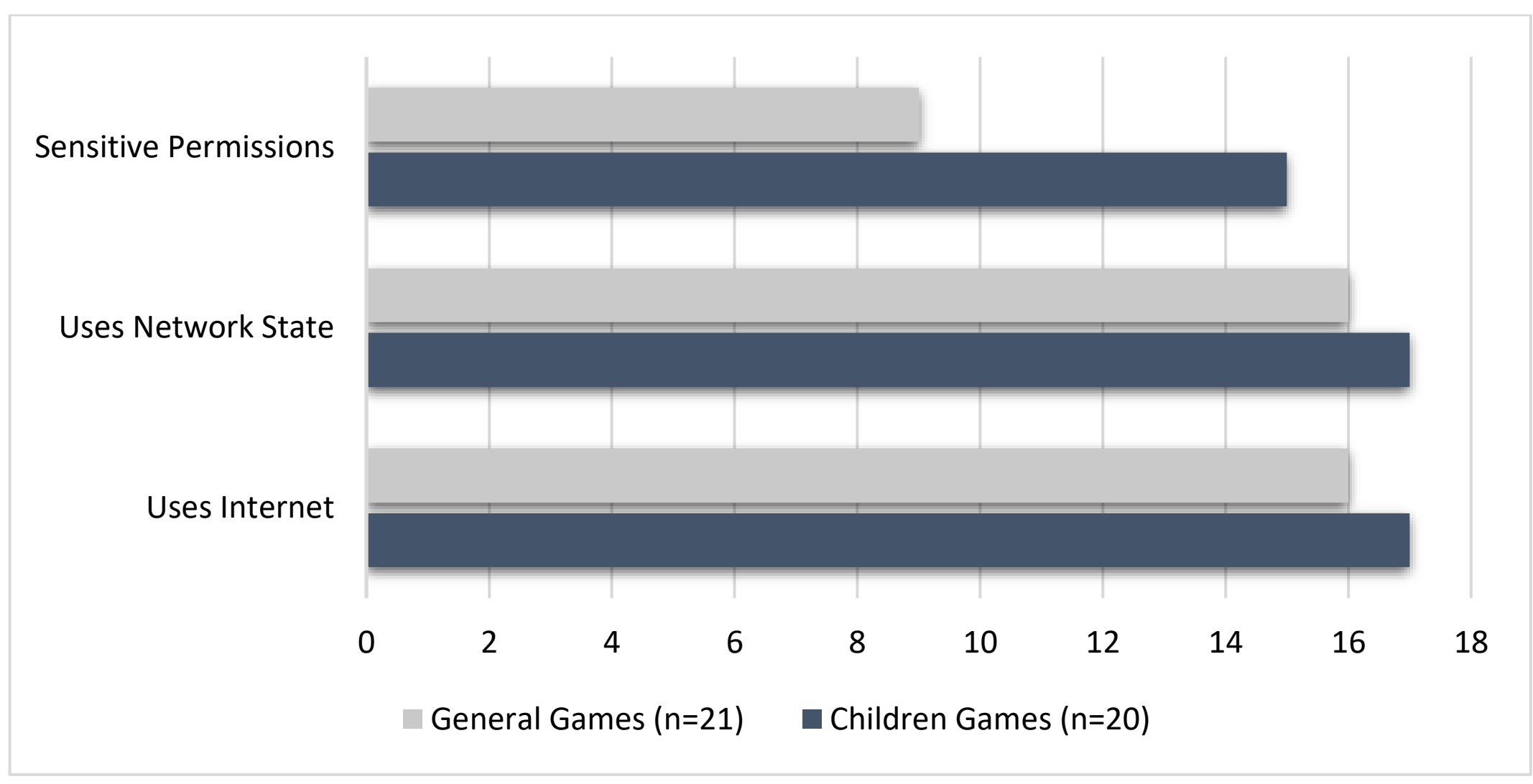

Figure 2. Comparison of Permission Usage Across Children-Oriented and General-Audience Mobile Games

Permission requests differ between children-oriented and general-audience games primarily in the presence of sensitive capabilities rather than total permission counts. While children-oriented apps generally rely on normal permissions associated with basic functionality, general-audience games more frequently request sensitive permissions that expand access to device resources. However, permission footprints alone provide only a partial view of privacy exposure. This observation motivates the need for deeper examination of configuration-level characteristics and third-party ecosystem dependencies, which may reveal additional privacy exposure indicators beyond permission declarations.

*4.2 Configuration-Level Security Indicators*

In addition to permission requests, the analysis examined configuration attributes defined within the Android app manifest, as described in Section 3. These configuration parameters influence how apps interact with the Android operating system and with other apps, and they may affect the potential exposure of locally stored data or application components. One configuration aspect considered in this study is the presence of exported app components, including activities and services that can be invoked by external apps. Exported components are commonly used to support inter-application communication, deep-linking functionality, and integration with external frameworks. However, prior research has shown that improperly exposed components may increase the potential attack surface of an application by allowing unintended interactions between apps [23].

Exported components are commonly used in Android applications to support inter-application communication, deep linking, and integration with external frameworks. However, prior research has shown that improperly exposed components may increase the attack surface of an application by allowing unintended interactions between applications, potentially leading to issues such as intent injection or component hijacking [32]. While these mechanisms are not inherently insecure, they require careful permission management and defensive programming practices. Across the analyzed dataset, exported activities were present in most applications. Children-oriented apps exposed an average of 2.55 exported activities, while general-audience games exposed approximately 2.90 activities on average. Exported services were also observed in several applications, with children-oriented games exposing approximately 1.20 services compared with 2.14 services in general-audience apps. Although the differences between categories are relatively modest, the presence of exported components across both groups indicates that inter-application communication mechanisms are widely used in mobile gaming apps. Figure 3 illustrates that exported components are common in both children-oriented and general apps, highlighting architectural patterns that may increase vulnerability to attacks. Examples in Figure 3 from App2 and App14 represent typical deep-linking and attribution exposure patterns observed.

```
<activity
  android:name="com.example.sdk.DeepLinkActivity"
   android:exported="true">
 <intent-filter>
 <action android:name="android.intent.action.VIEW" />
<category android:name="android.intent.category.DEFAULT" />
 <category
android:name="android.intent.category.BROWSABLE" />
<data android:scheme="https" android:host="example.com"/>
</intent-filter>
</activity>
```

```
<service
   android:name="com.example.analytics.UploadService"
   android:exported="true" />
```

Figure 3. Prevalence of Exported App Components Across the Dataset

Another configuration attribute examined in this study is the *android:allowBackup* setting, which determines whether application data may be included in Android backup mechanisms. When enabled, this configuration allows application data to be stored in device or cloud backup archives. Although backup functionality is not inherently insecure, it may expose locally stored data if backup files are accessed from compromised devices or insecure storage environments. Prior research has also highlighted that backup mechanisms can introduce unintended data exposure when applications store sensitive information locally [19]. Across the analyzed dataset, backup functionality was enabled in a substantial proportion of applications, indicating that backup-based data transfer pathways are common across both children-oriented and general-audience mobile games. Because this mechanism operates independently of runtime permission grants, locally stored artifacts (e.g., app preferences, cached telemetry, identifiers, or SDK-generated metadata, etc.) may be included in backup archives even when apps request relatively few sensitive permissions. As in Figure 4, a few examples from the dataset illustrate how backup configurations vary across apps. App9 explicitly disables backups (*allowBackup="false"*), demonstrating that conservative configurations are feasible, particularly for offline or self-contained games. In contrast, App7 enables backups and declares a custom *backupAgent* with permissive restore behavior (*restoreAnyVersion="true"*), which may broaden the scope of data restored across application versions. Similarly, App13 enables backups while referencing explicit backup and data-extraction rule files, suggesting selective control over backed-up content. However, the existence of backup pathways may still introduce potential data exposure depending on the completeness and correctness of such configuration rules. Overall, these observations suggest that configuration-level attributes contribute to privacy exposure alongside permission usage, reinforcing the need to examine app architecture and manifest settings when assessing privacy characteristics in mobile gaming apps.

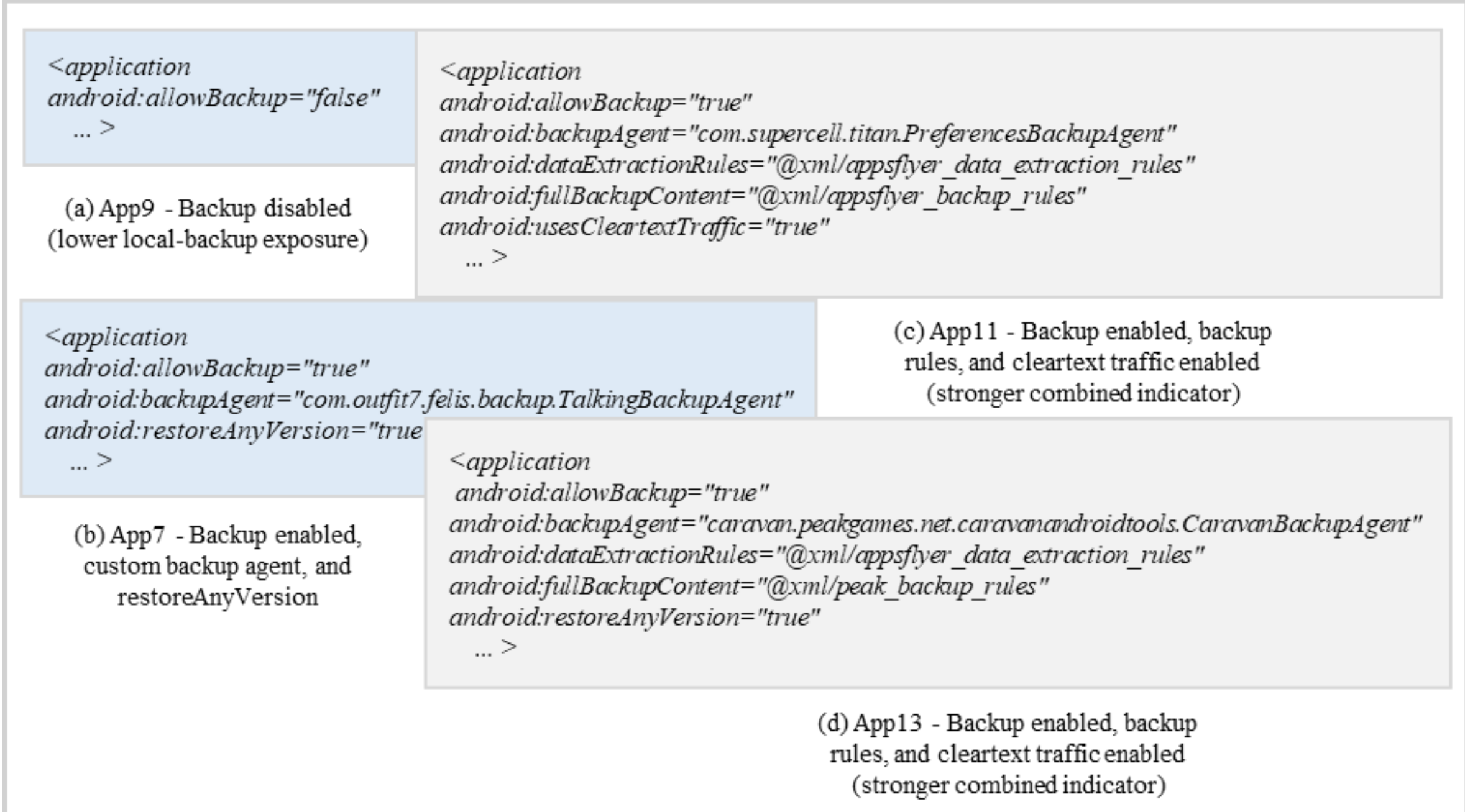

Figure 4. Backup Configuration and Local Data Exposure Indicators Across the Dataset

Configuration-level attributes may introduce privacy exposure independently of permission declarations. Backup-related settings can enable local data extraction through device backup mechanisms, while exported components—such as activities and services—were widely observed across the analyzed apps and were often associated with deep linking or third-party SDK integrations. These findings highlight how architectural configuration choices can influence privacy exposure beyond what is observable through permission requests alone.

*4.3 Network Communication Practices*

Secure communication between mobile apps and remote servers is critical for protecting user data during transmission. Prior research has demonstrated that insecure communication channels in Android apps may expose user data to interception or manipulation when encryption mechanisms are improperly implemented [33, 34]. Within the analyzed dataset, the *INTERNET* permission was present in 33 of the 41 apps (80.5%), while network state access was also observed in 33 apps (80.5%). These permissions are expected in modern mobile games because many applications rely on online services such as advertisements, analytics reporting, multiplayer functionality, and cloud-based updates.

Android apps can control transport security through the Network Security Configuration mechanism (*android:networkSecurityConfig*) and through the manifest attribute *android:usesCleartextTraffic*. When cleartext traffic is permitted, apps may allow unencrypted HTTP communication under certain conditions. Although this configuration does not necessarily indicate that sensitive data is transmitted without encryption, it may introduce a potential exposure condition if network communication occurs over unencrypted channels. Across the dataset, 8 apps (19.5%) permitted cleartext network communication, suggesting that unencrypted traffic may occur in some scenarios. For example, App2 and App11 declare both a network security configuration and explicit cleartext allowance, as illustrated in Figure 5(a). App14 specifies the network security configuration at both the manifest and application levels and explicitly enables cleartext traffic in the <*application*> element, as shown in Figure 5(b). In contrast, App7 references a network security configuration file (*android:networkSecurityConfig="@xml/network_security_config*"), where the allowance of cleartext traffic depends on the rules defined within the referenced configuration.

Similar observations have been reported in previous research, where Android applications transmitted sensitive information over insecure channels due to misconfigured network security settings [33-35]. Conversely, 14 applications (34.1%) included explicit network security configurations, indicating that some developers implemented additional controls to regulate network communication behavior. These configurations allow developers to define domain-level security policies and restrict cleartext communication when required. Overall, these findings suggest that network communication configurations may influence privacy exposure independently of permission declarations, reinforcing the importance of examining both manifest settings and runtime communication policies when evaluating the privacy characteristics of mobile apps.

```
android:networkSecurityConfig="@xml/network_security_config"
...
android:usesCleartextTraffic="true"
```

(a) App2 and App11 - Both apps declare a network security config and cleartext allowance

```
<manifest...
android:networkSecurityConfig="@xml/network_security_config"
...>
...
<application...
android:networkSecurityConfig="@xml/network_security_config"
...
android:usesCleartextTraffic="true">
```

(b) App14 -Network security config: manifest and app levels with cleartext enabled

Figure 5. Network Security Configuration and Cleartext Traffic Indicators

Several apps permit cleartext traffic or adopt permissive network security configurations that allow unencrypted communication channels. Although this does not necessarily indicate insecure data transmission, it may introduce potential exposure conditions independent of permission declarations. These findings highlight transport-layer security as an important configuration-level consideration in mobile apps.

### *4.4 Third-Party SDK Integration*

Third-party SDKs are widely used in mobile apps to support analytics, advertising, social integration, and development frameworks. However, prior research has shown that third-party libraries may introduce additional data collection practices and external communications beyond the direct control of app developers [17, 20]. The static analysis revealed extensive reliance on such SDKs across the analyzed dataset. In total, 32 of the 41 apps (78.0%) included Google-related SDKs, 21 apps (51.2%) integrated Unity-based development frameworks, 23 apps (56.1%) included advertising SDKs, and 22 apps (53.7%) incorporated social media SDKs. For example, App3, App4, and App5 were developed using Unity-based frameworks that embed Unity analytics and development services. Similarly, App10 and App13 integrated multiple external SDKs simultaneously, including advertising and analytics libraries. These integrations illustrate how modern mobile games frequently rely on complex third-party ecosystems to support functionality such as advertising attribution, user analytics, and social connectivity. As illustrated in Figure 6, third-party tracking and analytics infrastructures are particularly common in general-audience games, where several apps integrate multiple advertising and attribution frameworks. The examples presented in Figure 6 were extracted from the decoded AndroidManifest.xml files of App7, App11, and App13, and illustrate representative advertising, analytics, and attribution integration patterns observed across the dataset.

```
<meta-data
android:name="com.google.firebase.analytics.APPLICATION_ID"
   android:value="1:1234567890:android:abcdef"/>
<meta-data
 android:name="com.google.android.gms.ads.APPLICATION_ID"
   android:value="ca-app-pub-XXXXXXXXXXXXXXXX~YYYYYYYYYY" />
```

(a) Metadata declaring analytics and advertising SDKs

(b) Attribution / ad SDK receiver

```
Attribution / ad SDK receiver
<receiver
   android:name="com.appsflyer.SingleInstallBroadcastReceiver"
   android:exported="true">
   <intent-filter>
      <action
android:name="com.android.vending.INSTALL_REFERRER"/>
   </intent-filter>
</receiver>
```

Figure 6. Presence of Third-Party Analytics, Advertising, and Attribution SDKs

Importantly, the presence of third-party SDKs does not necessarily indicate privacy violations. However, these components may introduce additional data flows that are not always reflected in runtime permission requests. Many analytics and advertising frameworks rely primarily on network communication and application-level identifiers rather than sensitive device permissions. As a result, permission-based analysis alone may provide an incomplete view of privacy exposure, highlighting the importance of examining embedded SDK ecosystems alongside permission usage. Overall, the findings demonstrate that third-party SDK integrations represent a significant architectural factor shaping privacy exposure in mobile gaming apps.

Third-party analytics, advertising, and attribution SDKs are widespread across the analyzed apps, particularly in general-audience games. Although children-oriented apps also embed analytics components, they typically include fewer advertising dependencies. These observations indicate that ecosystem-level integrations may influence privacy exposure beyond what is visible through permission requests alone.

*4.5 Overall Privacy Exposure Indicators: Children-oriented Apps vs. General-audience Apps*

To synthesize the observations, each app was assigned an overall privacy exposure level based on the combination of indicators identified during the static inspection. The classification was derived from multiple indicators, including permission usage, backup configuration, network security settings, exported component exposure, and the integration of third-party SDKs. This categorization reflects potential exposure conditions inferred from static analysis rather than confirmed privacy violations and is intended to support comparative reporting across apps. The apps were categorized as high exposure when multiple strong indicators co-occurred. These indicators included extensive permission usage combined with permissive configuration settings (e.g., backup enablement or cleartext traffic allowances), the presence of numerous exported components, and the integration of multiple third-party SDKs such as analytics, advertising, or attribution frameworks. Medium exposure apps exhibited partial exposure conditions, such as moderate permission footprints, limited configuration-level indicators, or the presence of analytics or advertising infrastructures without additional high-impact configuration signals. Low exposure apps demonstrated comparatively conservative configurations, including minimal permission usage, restricted component exposure, limited external dependencies, and fewer configuration-level indicators associated with data access or network communication.

Across the 41 analyzed mobile gaming apps, 28 apps (68.3%) were classified as high exposure, 12 apps (29.3%) as medium exposure, and 1 app (2.4%) as low exposure, as in Table 1. For example, App3, App4, and App5 were categorized as high exposure due to the presence of multiple indicators, including extensive third-party SDK integration, multiple exported components, and active network connectivity with advertising and analytics infrastructures. In contrast, App9 was the only app categorized as low exposure, as it demonstrated a minimal permission footprint, limited external dependencies, and fewer configuration indicators associated with expanded data access or external service interactions. When comparing the exposure levels across categories, 14 of the 20 children-oriented apps (70.0%) and 14 of the 21 general-audience games (66.7%) were classified as high exposure.

Medium exposure was observed in 6 children-oriented apps (30.0%) and 6 general-audience games (28.6%), while one general-audience app (4.8%) was categorized as low exposure. These results indicate that elevated privacy exposure conditions are prevalent across both categories and suggest that privacy exposure in mobile gaming ecosystems is shaped primarily by architectural configurations and third-party ecosystem integrations rather than the intended user audience alone. Overall, the findings indicate that the majority of the analyzed mobile gaming apps exhibit multiple architectural indicators associated with elevated privacy exposure. This pattern suggests that privacy exposure is driven not only by permission requests but also by broader architectural factors, including configuration-level decisions and the integration of third-party analytics, advertising, and attribution frameworks. These results highlight the importance of privacy-aware design practices in mobile gaming ecosystems where external services and network-based functionality are extensively integrated.

Collectively, these findings provide empirical evidence that privacy exposure emerges from the interaction of configuration-level design decisions and third-party ecosystem dependencies, rather than from permission usage alone. This underscores the limitations of single-dimensional assessment approaches and reinforces the need for multi-dimensional, configuration-aware methodologies for evaluating privacy exposure in modern mobile software systems.

Table 1. Distribution of Privacy Exposure Indicators Across App Categories.

| Exposure *Assessment* | High Exposure | Medium Exposure | Low Exposure | Total |
|---|---|---|---|---|
| *Children-oriented* | 14 (70.0%) | 6 (30.0%) | 0 (0%) | 21 |
| *General-audience* | 14 (66.7%) | 6 (28.6%) | 1 (4.8%) | 20 |
| *Total* | 28 (68.3%) | 12 (29.3%) | 1 (2.4%) | 41 |

## 5. Discussion of the Key Findings

This study investigated privacy and security exposure indicators in 41 Android mobile gaming apps, including both children-oriented and general-audience games. The results indicate that privacy exposure in mobile games emerges from multiple architectural factors, including permission usage, configuration-level design decisions, network communication practices, and the integration of third-party SDKs. Importantly, the indicators reported in this study represent potential exposure pathways derived from static analysis, rather than verified vulnerabilities or confirmed data misuse. Static inspection techniques have been widely used in prior research to analyze privacy and security exposures in mobile apps [36-38]. Recent research continues to adopt similar empirical approaches to understand how app architecture and third-party ecosystems influence privacy outcomes [21, 22]. The following subsections interpret the findings in relation to previous studies and discuss their implications for software engineers, platform providers, and users. Furthermore, we provide a consolidated view of actionable guidance, evidence-driven recommendations derived from the observed exposures, which are summarized in Table 2.

### *5.1 Permission Usage and Privacy Exposure*

The results show that general-audience games request significantly more permissions than children-oriented games, with averages of 17.43 permissions per app compared with 11.35 permissions per app in children-oriented titles. Additionally, 68.3 % of the analyzed apps requested at least one permission categorized as sensitive. These findings align with previous research demonstrating that many Android apps request permissions beyond those strictly required for their core functionality. Felt et al. in [15] showed that over-privileged permission requests are common in Android apps, while more recent research indicates that permission usage alone does not necessarily reflect actual data-collection behavior as in [39]. More importantly, the results of this study support the growing body of research suggesting that privacy exposure cannot be accurately evaluated based solely on permission prompts. Even apps with relatively limited permission sets may still collect data through third-party libraries or background services. Previous empirical analyses of Android ecosystems have similarly found that third-party SDKs frequently access sensitive information through indirect mechanisms that are not always visible through the permission model [17, 20]. Therefore, while permission minimization remains a fundamental security principle, developers should also examine how permissions interact with external libraries and network communication mechanisms when assessing privacy exposure.

### *5.2 Configuration-Level Design Decisions*

Beyond permission usage, the findings highlight the importance of configuration-level attributes defined in Android app manifests, including exported components and backup settings. Several apps in the dataset contained exported activities or services that could potentially be invoked by other apps. Prior research has shown that improperly exposed components may allow unintended inter-application interactions, increasing the potential attack surface of mobile apps [23]. Another configuration examined in this study was the *android:allowBackup* attribute, which determines whether application data may be included in Android backup mechanisms. Enabling backup functionality does not inherently constitute a security vulnerability, as backup features are commonly used to support data portability and device migration. However, previous research has shown that backup mechanisms can expose locally stored application data when sensitive files are included in backup archives or when backup data can be extracted through forensic or debugging tools [40]. In addition, Android security guidelines emphasize that improper backup configurations may lead to unintended data disclosure if sensitive information is not explicitly excluded from backup processes [41]. Accordingly, this study treats backup configuration as a potential privacy exposure indicator rather than a confirmed vulnerability. Recent work on Android software supply chains further emphasizes that privacy exposure often arises from configuration decisions combined with third-party SDK integration rather than from a single architectural element. For example, recent analyses of Android SDK ecosystems show that configuration settings and default behaviors can significantly affect data-collection practices [22]. These observations suggest that configuration-level decisions represent an important but often overlooked dimension of privacy-aware software engineering.

### *5.3 Network Communication and Secure Data Transmission*

Secure network communication is essential for protecting user data transmitted between mobile applications and remote services. The analysis showed that 80.5 % of the studied apps requested internet access, which reflects the heavy reliance of mobile gaming applications on online connectivity for analytics, advertisements, and content updates. However, 19.5 % of the apps permitted cleartext network traffic, indicating that unencrypted communication could occur under certain conditions. While this configuration does not necessarily imply that sensitive data is transmitted insecurely, it represents a condition that may increase exposure to interception attacks. Previous research has documented similar concerns regarding insecure communication practices in mobile applications. Fahl et al. in [33] demonstrated that many Android applications historically transmitted sensitive information through insecure channels due to incorrect implementations of SSL/TLS certificate validation. More recent research suggests that insecure network communication and weak encryption practices remain relevant challenges in mobile software ecosystems. For example, Yang et al. in [42] identified widespread TLS misconfigurations in Android applications that could enable Man-in-the-Middle attacks, while Forsberg and Iwaya in [25] reported insecure cryptographic implementations and extensive communication with external domains in widely deployed mobile applications. Together, these findings highlight the importance of adopting secure-by-default communication policies, particularly for apps that integrate external services and analytics infrastructures.

### *5.4 Third-Party SDK Ecosystems*

One of the most significant findings of this study is the widespread reliance on third-party SDKs and external development frameworks. The dataset revealed that 78 % of the analyzed apps integrated Google-related SDKs, while more than 50 % incorporated Unity frameworks, advertising SDKs, or social media libraries. These observations align with previous research highlighting the central role of third-party libraries in mobile app ecosystems. Earlier work demonstrated that advertising libraries embedded within Android apps frequently collect device identifiers and behavioral data [20]. More recent empirical research confirms that SDK ecosystems remain a major source of privacy exposure in mobile applications. A large-scale study by Meng et al. in [21] analyzed 158 widely used Android SDKs and identified hundreds of instances of privacy-sensitive data flows and widespread inconsistencies in privacy disclosures. Similarly, recent research by Rodriguez et al. in [22] examined privacy configuration options in third-party SDKs and found that developers often rely on default library settings, which may enable additional data collection or tracking behaviors unless explicitly disabled. These findings highlight the central role that SDK ecosystems play in shaping privacy exposure in modern mobile apps.

### *5.4 Privacy Challenges in Children-Oriented Mobile Games*

A key finding of this study is that children-oriented mobile games are not necessarily characterized by lower privacy exposure. Although several children-oriented games request fewer permissions than general-audience games, many still exhibited potential privacy exposure indicators arising from configuration-level settings, embedded tracking infrastructures, and the integration of third-party SDKs. In fact, the comparative analysis showed that 70% of the analyzed children-oriented apps and 66.7% of the general-audience apps were classified as high exposure, indicating that elevated exposure conditions are prevalent across both categories. This finding challenges the common assumption that age-based labeling or educational positioning reliably signals privacy-preserving design practices. Prior research has similarly reported that apps targeting children may still incorporate advertising or analytics infrastructures capable of collecting behavioral data or communicating with external services, despite being marketed as child-friendly applications [43]. These observations suggest that compliance claims or target audience labels do not always correspond to conservative technical implementations in mobile apps. Instead, the results indicate that privacy exposure in mobile games is shaped by a combination of architectural design decisions, monetization strategies, and dependencies on third-party ecosystems, rather than by audience designation alone. Consequently, evaluating the privacy characteristics of mobile gaming apps requires examining not only permission usage but also configuration-level settings, component exposure, and external service integrations that may influence how user data is collected, processed, and transmitted.

### *5.5 Implications for Software Engineering Practices*

The findings of this study highlight several implications for software engineering practice. First, configuration-level design decisions, such as backup policies, network security configurations, and component exposure, should be treated as first-class privacy design concerns rather than secondary implementation details. Second, the widespread integration of third-party SDKs suggests that privacy exposure is often influenced by external ecosystem dependencies, requiring developers to carefully evaluate SDK selection, configuration, and default behaviors. Third, the results indicate that privacy exposure should be assessed as an emergent architectural property rather than through isolated indicators such as permission usage alone. These insights support the need for incorporating configuration-aware privacy assessment techniques into software development and evaluation processes.

### *5.6 Implications for Software Engineers and End-Users*

The findings of this study provide several practical implications for both software engineers and end-users. In the context of children-oriented mobile apps, end-users may include parents or guardians who install, manage, or supervise the use of mobile games on children's devices. Consequently, improving privacy outcomes in mobile gaming ecosystems requires awareness and coordinated efforts from both developers and those responsible for selecting and managing apps for young users. For software engineers, the results highlight the importance of adopting privacy-by-design principles throughout the software development lifecycle. Privacy-by-design emphasizes integrating privacy protections into application architecture and development processes from the earliest stages rather than addressing privacy only during compliance or deployment phases [44] . In practice, this includes minimizing permission requests, restricting unnecessary exported components, enforcing encrypted communication channels, and carefully evaluating and auditing third-party SDK integrations. Developers should also consider reducing dependencies on advertising or analytics infrastructures in apps targeting children, as these components may introduce additional data collection pathways that are not immediately visible to end-users.

For end-users, including parents and guardians managing children's devices, the findings indicate that app age ratings or educational labels should not be interpreted as guarantees of privacy-preserving design. Even apps marketed for children may integrate analytics or advertising frameworks capable of collecting behavioral information. As a result, parents and guardians should consider additional indicators when selecting mobile games for children, such as the presence of advertising, transparency of privacy policies, and the reputation of the app publisher. Prior research on mobile privacy perception has similarly shown that users and caregivers often underestimate the extent of data collection performed by mobile apps [29]. To support these practical implications, Table 2 summarizes key privacy exposure indicators identified in this study and maps them to actionable recommendations for both end-users and developers.

Table 2. Evidence-Driven Privacy and Security Recommendations for Mobile Apps for Gaming

| *Issue Identified* | *Observed Evidence* | *Implications* | *Recommendations for End-Users* | *Recommendations for Developers* |
|---|---|---|---|---|
| Permission footprint limitations | Apps with few permissions still embed analytics and advertising SDKs. | Permissions requests alone underestimate potential privacy exposure. | - Do not rely solely on permission prompts.<br>- Prefer transparent publishers with clear privacy disclosures. | - Disclose tracking clearly.<br>- Minimize reliance on third-party SDKs where possible |
| Backup and local data exposure | The configuration attribute *android:allowBackup ="true" was* present in multiple analyzed apps. | App data may be extractable via device backups if not properly configured. | - Enable device encryption.<br>- Use secure device lock mechanisms. | - Disable backups by when unnecessary.<br>- Exclude sensitive data via rules backup configuration rules. |

| Cleartext network communication | Cleartext allowances observed in network security configurations. | Unencrypted communication may increase the exposure of data interception in insecure network environments. | - Avoid public Wi-Fi networks for gaming apps. | - Enforce TLS by default.<br>- Restrict cleartext to trusted domains if required. |
|---|---|---|---|---|
| Exported application components | Several apps contained exported activities or services without additional protection mechanisms. | Exported components may increase the application attack surface through unintended inter-application communication. | - Limit installation of unknown or sideloaded apps on children's devices. | - Restrict exported components when possible Enforce permissions.<br>- Apply explicit permission protection to exported components. |
| Third-party SDK proliferation | Many apps integrated multiple advertising and analytics SDKs from different vendors. | Multiple SDK dependencies may increase data sharing complexity and reduce transparency regarding data flows. | - Minimize app installations on children's devices.<br>- Review privacy policies before installation. | - Reduce the number of embedded SDKs.<br>- Audit third-party libraries and associated data flows. |
| Privacy exposure in children-oriented apps | Several children-oriented apps exhibited medium levels of potential privacy exposure despite relatively conservative permission usage | Age-based labels or child-oriented positioning do not necessarily guarantee privacy-preserving design. | - Prefer child profiles and parental controls.<br>- Favor apps without advertising infrastructures. | - Apply privacy-by-design practices when developing child-directed apps.<br>- Avoid unnecessary advertising or tracking SDKs. |

## 6. Threat to Validity

As with any empirical study, several limitations may impact the validity of the findings. Following common practice in empirical software engineering research, we discuss potential threats related to internal, construct, external, and conclusion validity, along with mitigation strategies, which will be discussed below.

### *6.1 Internal Validity*

Internal validity may be affected by the reliance on static analysis techniques, which analyze application artifacts without executing the apps. Static inspection may not capture behaviors that occur only during runtime, such as remote configuration, dynamic code loading, or interactions triggered by specific user actions or device conditions. In addition, some functionalities of embedded SDKs may remain dormant unless particular runtime conditions are met. To mitigate this limitation, the analysis relied on multiple complementary indicators, including permissions, manifest configuration attributes, exported component exposure, and evidence of third-party SDK integration. By triangulating these indicators, the study reduces dependence on a single signal and provides a broader view of potential privacy exposure pathways.

### *6.2 Construct Validity*

Construct validity concerns whether the selected indicators appropriately represent privacy exposure. In this study, privacy exposure is operationalized using observable technical signals such as permission usage, cleartext traffic configuration, backup settings, and the presence of tracking or advertising SDKs. These indicators do not directly measure actual data collection or misuse. Instead, they represent architectural conditions that may increase potential privacy exposure. To address this limitation, the study interprets the indicators conservatively and avoids claims of confirmed privacy violations. The same operational criteria were applied consistently across all analyzed apps.

### *6.3 External Validity*

External validity is limited by the scope of the dataset. The study analyzed 41 widely used Android mobile gaming apps, including both children-oriented and general-audience titles. While this dataset provides useful insights into common development practices, it may not fully represent the broader diversity of the Android gaming ecosystem, including niche or region-specific apps. To mitigate this limitation, the study focused on widely deployed apps and documented a repeatable analysis procedure so that future studies can replicate or extend the dataset with additional apps or even with newer versions.

*6.4 Conclusion Validity*

Conclusion validity may be affected by the qualitative categorization of privacy exposure levels (e.g., low, medium, high), which can introduce some subjectivity. To reduce this exposure, the categorization was based on clearly defined combinations of technical indicators rather than subjective judgment. In addition, the study avoids causal claims about privacy violations and restricts conclusions to observations supported by the available evidence.

*6.5 Summary*

Despite the limitations associated with static analysis and dataset scope, the use of multiple indicators, transparent operational rules, and conservative interpretation helps strengthen confidence in the reported findings and supports their use in future research.

## 7. Conclusions and Future Work

Mobile gaming apps have become one of the most widely adopted categories of mobile software, attracting a diverse range of users including children and young audiences. Despite their popularity, the privacy implications of mobile gaming apps remain insufficiently understood, particularly with respect to architectural design choices that may expose user data beyond the visibility of traditional permission mechanisms. This study presented an empirical analysis of 41 Android mobile gaming apps, including both children-oriented and general-audience games. Through a structured static inspection of permissions, configuration attributes, network communication settings, and third-party SDK integrations, the study investigated architectural indicators that may influence privacy exposure in mobile gaming apps. The results show that privacy exposure in mobile gaming apps cannot be assessed solely through permission requests. Instead, exposure often arises from the interaction between permissions, configuration decisions, network communication practices, and third-party SDK ecosystems. The analysis revealed several important observations and key findings:

- Permission usage alone does not fully represent privacy exposure. While general-audience games requested more permissions on average than children-oriented games, some apps with relatively limited permission sets still exhibited architectural indicators associated with increased privacy exposure.
- Configuration-level design decisions play an important role. Settings such as exported components or backup configurations can influence how app functionality and data are exposed within the Android ecosystem.
- Network communication practices remain a relevant factor. A portion of the analyzed apps permitted cleartext communication, indicating that secure network configurations are not consistently enforced across mobile gaming applications.
- Third-party SDK ecosystems significantly shape privacy exposure. Many analyzed apps relied on multiple external libraries for analytics, advertising, and game development frameworks, which may introduce additional data flows beyond the direct control of the app developer.

Together, these findings reinforce the importance of considering multiple architectural dimensions when evaluating privacy exposure in mobile apps. We believe this study contributes to the existing body of research on mobile app privacy in several ways:

- *Empirical analysis of mobile gaming apps:* The study provides a structured examination of privacy exposure indicators across 41 widely used Android gaming apps, including both children-oriented and general-audience titles.
- *Architectural perspective on privacy exposure:* Rather than focusing solely on permission requests, the study examines a broader set of architectural elements, including configuration settings, network communication practices, and SDK dependencies.
- *Insights for privacy-aware software engineering:* The findings highlight how development decisions related to permissions, configurations, and external libraries can collectively influence privacy exposure in mobile apps.
- *Implications for child-directed applications:* The study demonstrates that children-oriented apps should not automatically be assumed to be privacy-preserving simply because they request fewer permissions or are labelled as suitable for children.

From a practical standpoint, the results emphasize the importance of adopting privacy-by-design principles during mobile app development. Developers should carefully evaluate permission requirements, enforce secure communication practices, restrict unnecessary component exposure, and critically assess the necessity and configuration of third-party SDK integrations. The findings of this study open several opportunities for future research such as:

- *Dynamic analysis of mobile apps:* Future studies could complement static inspection with runtime analysis techniques to observe how mobile gaming apps transmit data and how third-party SDKs behave during actual execution.
- *Larger-scale empirical studies:* Expanding the dataset to include a broader range of apps may help identify privacy exposure patterns across different mobile software categories and development frameworks.
- *Developer-centered investigations:* Qualitative studies involving interviews with mobile apps developers could provide deeper insights into how privacy considerations influence architectural and design decisions during mobile apps development.
- *Platform governance and policy evaluation:* Future research could examine how platform policies, developer guidelines, and privacy regulations influence privacy practices in child-directed mobile apps.

Overall, this study contributes to the growing body of research on mobile privacy by providing empirical evidence that privacy exposure in mobile gaming apps is shaped by multiple interacting architectural factors. By highlighting these relationships, the study aims to support more privacy-aware software engineering practices and encourage further investigation into privacy protection in mobile software ecosystems.

## APPENDIX

The list of mobile gaming apps used in our study and the full results for each app can be found in the following links https://shorturl.at/93lrm

## ACKNOWLEDGMENTS

The author gratefully acknowledged the valuable comments and suggestions provided by anonymous reviewers on an earlier version of this manuscript, which significantly contributed to improving its quality. The author also thanks Akram Aloafi for sharing his input during the early stages of this study.

## References

[1] "App downloads by category in 2025 available at https://www.apptweak.com/en/reports/app-downloads-by-category," 2025.

[2] "Mobile Gaming Market (2025 - 2030) avaliable at https://www.grandviewresearch.com/industry-analysis/mobile-games-market."

[3] R. Sun, M. Xue, G. Tyson, S. Wang, S. Camtepe, and S. Nepal, "Not seen, not heard in the digital world! measuring privacy practices in children’s apps," in *Proceedings of the ACM Web Conference 2023*, 2023, pp. 2166-2177.

[4] R. Carlsson, S. Rauti, S. Laato, T. Heino, and V. Leppänen, "Privacy in popular children’s mobile applications: A network traffic analysis," in *2023 46th MIPRO ICT and Electronics Convention (MIPRO)*, 2023, pp. 1213-1218.

[5] P. Laperdrix, N. Mehanna, A. Durey, and W. Rudametkin, "The price to play: A privacy analysis of free and paid games in the android ecosystem," in *Proceedings of the ACM Web Conference 2022*, 2022, pp. 3440-3449.

[6] A. Razaghpanah, R. Nithyanand, N. Vallina-Rodriguez, S. Sundaresan, M. Allman, C. Kreibich*, et al.*, "Apps, trackers, privacy, and regulators: A global study of the mobile tracking ecosystem," in *The 25th annual network and distributed system security symposium (NDSS 2018)*, 2018.

[7] R. Binns, U. Lyngs, M. Van Kleek, J. Zhao, T. Libert, and N. Shadbolt, "Third party tracking in the mobile ecosystem," in *Proceedings of the 10th ACM Conference on Web Science*, 2018, pp. 23-31.

[8] F. Paci, J. Pizzoli, and N. Zannone, "A comprehensive study on third-party user tracking in mobile applications," in *Proceedings of the 18th international conference on availability, reliability and security*, 2023, pp. 1-8.

[9] Y. Yang, M. Elsabagh, C. Zuo, R. Johnson, A. Stavrou, and Z. Lin, "Detecting and measuring misconfigured manifests in android apps," in *Proceedings of the 2022 ACM SIGSAC Conference on Computer and Communications Security*, 2022, pp. 3063-3077.

[10] A. K. Jha, S. Lee, and W. J. Lee, "Developer mistakes in writing android manifests: An empirical study of configuration errors," in *2017 IEEE/ACM 14th International Conference on Mining Software Repositories (MSR)*, 2017, pp. 25-36.

[11] R. Carlsson, S. Rauti, S. Laato, T. Heino, and V. Leppänen, "Privacy in Popular Children's Mobile Applications: A Network Traffic Analysis," in *MIPRO ICT and Electronics Convention*, 2023, pp. 1213-1218.

[12] A. Nairn and D. Monkgol, "Children and privacy online," *Journal of Direct, Data and Digital Marketing Practice,* vol. 8, pp. 294-308, 2007.

[13] I. Reyes, P. Wijesekera, J. Reardon, A. Elazari Bar On, A. Razaghpanah, N. Vallina-Rodriguez*, et al.*, "“Won't somebody think of the children?” examining COPPA compliance at scale," in *The 18th Privacy Enhancing Technologies Symposium (PETS 2018)*, 2018.

[14] K. Langton and R. Ng, "“Tracking the Trackers” of children's first personal data in mobile applications: Using static analysis and privacy policy evaluation to explore the data-sharing capabilities and practices of baby apps," in *Proceedings of the 37th Australian Conference on Human-Computer Interaction*, 2025, pp. 845-859.

[15] A. P. Felt, E. Ha, S. Egelman, A. Haney, E. Chin, and D. Wagner, "Android permissions: User attention, comprehension, and behavior," in *Proceedings of the eighth symposium on usable privacy and security*, 2012, pp. 1-14.

[16] A. P. Felt, S. Egelman, M. Finifter, D. Akhawe, and D. A. Wagner, "How to Ask for Permission," *HotSec,* vol. 12, pp. 7-7, 2012.

[17] J. Reardon, Á. Feal, P. Wijesekera, A. E. B. On, N. Vallina-Rodriguez, and S. Egelman, "50 ways to leak your data: An exploration of apps' circumvention of the android permissions system," in *28th USENIX security symposium (USENIX security 19)*, 2019, pp. 603-620.

[18] S. Zimmeck, Z. Wang, L. Zou, R. Iyengar, B. Liu, F. Schaub*, et al.*, "Automated Analysis of Privacy Requirements for Mobile Apps," in *NDSS*, 2017, pp. 1.4-2.3.

[19] L. Li, T. F. Bissyandé, M. Papadakis, S. Rasthofer, A. Bartel, D. Octeau*, et al.*, "Static analysis of android apps: A systematic literature review," *Information and Software Technology,* vol. 88, pp. 67-95, 2017.

[20] M. C. Grace, W. Zhou, X. Jiang, and A.-R. Sadeghi, "Unsafe exposure analysis of mobile in-app advertisements," in *Proceedings of the fifth ACM conference on Security and Privacy in Wireless and Mobile Networks*, 2012, pp. 101-112.

[21] M. H. Meng, C. Yan, Y. Hao, Q. Zhang, Z. Wang, K. Wang*, et al.*, "A large-scale privacy assessment of Android third-party sdks," *arXiv preprint arXiv:2409.10411,* 2024.

[22] D. Rodriguez, J. A. Calandrino, J. M. Del Alamo, and N. Sadeh, "Privacy settings of third-party libraries in android apps: A study of facebook sdks," *Proceedings on Privacy Enhancing Technologies,* 2025.

[23] E. Chin, A. P. Felt, K. Greenwood, and D. Wagner, "Analyzing inter-application communication in Android," in *Proceedings of the 9th international conference on Mobile systems, applications, and services*, 2011, pp. 239-252.

[24] A. Alkinoon, T. C. Dang, A. Alghuried, A. Alghamdi, S. Choi, M. Mohaisen*, et al.*, "A comprehensive analysis of evolving permission usage in Android apps: trends, threats, and ecosystem insights," *Journal of Cybersecurity and Privacy,* vol. 5, p. 58, 2025.

[25] A. Forsberg and L. H. Iwaya, "Security analysis of top-ranked mhealth fitness apps: an empirical study," in *Nordic Conference on Secure IT Systems*, 2024, pp. 364-381.

[26] S. Chen, T. Su, L. Fan, G. Meng, M. Xue, Y. Liu*, et al.*, "Are mobile banking apps secure? what can be improved?," in *Proceedings of the 2018 26th ACM Joint Meeting on European Software Engineering Conference and Symposium on the Foundations of Software Engineering*, 2018, pp. 797-802.

[27] N. Alomar and S. Egelman, "Developers say the darnedest things: Privacy compliance processes followed by developers of child-directed apps," *Proceedings on Privacy Enhancing Technologies,* 2022.

[28] N. Alomar, J. Reardon, A. Girish, N. Vallina-Rodriguez, and S. Egelman, "The effect of platform policies on app privacy compliance: A study of child-directed apps," *Proceedings on Privacy Enhancing Technologies 2025,* 2025.

[29] S. Saeed, "Usable privacy and security in mobile applications: perception of mobile end users in Saudi Arabia," *Big Data and Cognitive Computing,* vol. 8, p. 162, 2024.

[30] "OWASP Mobile Top 10 available at https://owasp.org/www-project-mobile-top-10/," 2023.

[31] K. W. Y. Au, Y. F. Zhou, Z. Huang, and D. Lie, "Pscout: analyzing the android permission specification," in *Proceedings of the 2012 ACM conference on Computer and communications security*, 2012, pp. 217-228.

[32] K. Lian, L. Zhang, G. Yang, S. Mao, X. Wang, Y. Zhang*, et al.*, "Component security ten years later: An empirical study of cross-layer threats in real-world mobile applications," *Proceedings of the ACM on Software Engineering,* vol. 1, pp. 70-91, 2024.

[33] S. Fahl, M. Harbach, T. Muders, L. Baumgärtner, B. Freisleben, and M. Smith, "Why Eve and Mallory love Android: An analysis of Android SSL (in) security," in *Proceedings of the 2012 ACM conference on Computer and communications security*, 2012, pp. 50-61.

[34] J. Ren, A. Rao, M. Lindorfer, A. Legout, and D. Choffnes, "Recon: Revealing and controlling pii leaks in mobile network traffic," in *Proceedings of the 14th Annual International Conference on Mobile Systems, Applications, and Services*, 2016, pp. 361-374.

[35] M. Georgiev, S. Iyengar, S. Jana, R. Anubhai, D. Boneh, and V. Shmatikov, "The most dangerous code in the world: validating SSL certificates in non-browser software," in *Proceedings of the 2012 ACM conference on Computer and communications security*, 2012, pp. 38-49.

[36] W. Enck, P. Gilbert, S. Han, V. Tendulkar, B.-G. Chun, L. P. Cox*, et al.*, "Taintdroid: an information-flow tracking system for realtime privacy monitoring on smartphones," *ACM Transactions on Computer Systems (TOCS),* vol. 32, pp. 1-29, 2014.

[37] Y. Pan, X. Ge, C. Fang, and Y. Fan, "A systematic literature review of android malware detection using static analysis," *Ieee Access,* vol. 8, pp. 116363-116379, 2020.

[38] L. Yu, X. Luo, X. Liu, and T. Zhang, "Can we trust the privacy policies of android apps?," in *2016 46th Annual IEEE/IFIP International Conference on Dependable Systems and Networks (DSN)*, 2016, pp. 538-549.

[39] K. Heid, E. J. Sonntag, and J. Heider, "Revisiting Permission Piggybacking of Third-Party Libraries in Android Apps," in *ICISSP (1)*, 2025, pp. 39-46.

[40] U. Hur, M. Park, G. Kim, Y. Park, I. Lee, and J. Kim, "Data acquisition methods using backup data decryption of Sony smartphones," *Digital Investigation,* vol. 31, p. 200890, 2019.

[41] P. Dutson, *Android development patterns: best practices for professional developers*: Addison-Wesley Professional, 2016.

[42] H. Yang, R. Huang, Y. Fang, B. Zhang, J. Guo, Z. Wu*, et al.*, "Okara: Detection and Attribution of TLS Man-in-the-Middle Vulnerabilities in Android Apps with Foundation Models," *arXiv preprint arXiv:2601.22770,* 2026.

[43] N. Alomar, J. Reardon, A. Girish, N. Vallina-Rodriguez, and S. Egelman, "The Effect of Platform Policies on App Privacy Compliance: A Study of Child-Directed Apps," *Proceedings on Privacy Enhancing Technologies,* 2025.

[44] A. Cavoukian, "Privacy by design: The 7 foundational principles," *Information and privacy commissioner of Ontario, Canada,* vol. 5, p. 12, 2009.